\documentclass{article}
\usepackage{amsmath}

\numberwithin{equation}{section}
\input{tcilatex}
\begin{document}

\centerline{\large\bf  Relationship between the n-tangle and the
residual entanglement of even n qubits\footnote{The paper was
supported by NSFC(Grants No.
 10875061,60433050, and 60673034 ) and Tsinghua National Laboratory for Information Science and Technology. }}

\centerline{Xiangrong Li$^{a}$, Dafa Li$^{b}$ }

\centerline{$^a$ Department of Mathematics, University of
California, Irvine, CA 92697-3875, USA}

\centerline{$^b$ Dept of mathematical sciences, Tsinghua University,
Beijing 100084 CHINA}

Abstract

We show that $n$-tangle, the generalization of the 3-tangle to even $n$
qubits, is the square of the SLOCC polynomial invariant of degree 2. We find
that the $n$-tangle is not the residual entanglement for any even $n\geq 4$\
qubits. We give a necessary and sufficient condition for the vanishing of
the concurrence $C_{1(2...n)}$. The condition implies that the concurrence $%
C_{1(2...n)}$ is always positive for any entangled states while the $n$%
-tangle vanishes for some entangled states. We argue that for even $n$\
qubits, the concurrence $C_{1(2...n)}$\ is equal to or greater than the $n$%
-tangle. Further,\ we reveal that the residual entanglement is a partial
measure for product states of any $n$ qubits while the $n$-tangle is
multiplicative for some product states.

Keywords: the 3-tangle, the $n$-tangle of even $n$ qubits, the residual
entanglement, SLOCC polynomial invariants

PACS numbers: 03.67.Mn, 03.65.Ud

\twocolumn

\section{Introduction}

Quantum entanglement is an important physical resource in quantum
information and computation such as quantum teleportation, cloning and
encryption. Entanglement phenomenon distinguishes the quantum world from the
classical world. Considerable attention has been paid in recent years to the
quantification and classification of entanglement. The concurrence was
proposed by Wootters in 1998 to quantify entanglement for bipartite systems 
\cite{Wootters}. For two qubits, the concurrence was defined as $%
C_{12}=Max\{0,\lambda _{1}-\lambda _{2}-\lambda _{3}-\lambda _{4}\}$, where $%
\lambda _{i}^{2}$ are the eigenvalues, in decreasing order, of $\rho _{12}%
\tilde{\rho}_{12}$. Here, $\rho _{12}$ is the density matrix and $\tilde{\rho%
}_{12}$ is the \textquotedblleft spin-flipped\textquotedblright density
matrix of $\rho _{12}$, i.e., $\tilde{\rho}_{12}=\sigma _{y}\otimes \sigma
_{y}$ $\rho _{12}^{\ast }\sigma _{y}\otimes \sigma _{y}$ \cite{Coffman},
where the asterisk denotes complex conjugation in the standard basis.\ For
the state $|\psi \rangle $ of a bipartite system, the concurrence was also
given by \cite{Rungta} 
\begin{equation}
C(\psi )=\sqrt{2(1-Tr(\rho _{A}^{2}))}.  \label{concur-def}
\end{equation}%
The definition of the concurrence in Eq. (\ref{concur-def}) was generalized
to multipartite systems \cite{Aolita}. Recently, the concurrence was used to
study quantum phase transitions \cite{Osterloh-nature}.

By means of the concurrence, CKW monogamy inequality for three qubits was
established. Namely, $C_{12}^{2}+C_{13}^{2}\leq C_{1(23)}^{2}$ \cite{Coffman}%
. Here $\rho _{12}$ is obtained from the density matrix $\rho _{123}$ by
tracing out over qubit 3, and $C_{1(23)}^{2}=4\det \rho _{1}$, where $\rho
_{1}=tr_{23}\rho _{123}$. Note that $C_{1(23)}$ can be called the
concurrence between qubit 1 and the pair of qubits 2 and 3 if qubits 2 and 3
are regarded as a single object. The difference $%
(C_{1(23)}^{2}-(C_{12}^{2}+C_{13}^{2}))$\ between the two sides of the above
CKW monogamy inequality is called \textquotedblleft residual
entanglement\textquotedblright . The algebraic expression for the residual
entanglement is called the 3-tangle (see (20) of \cite{Coffman} for the
expression). The expression can also be obtained from Eq. (\ref{n-tangle-1})
by letting $n=3$. The 3-tangle is invariant under permutations of all the
qubits \cite{Coffman}. The invariance of entanglement measure under
permutations of all the qubits represents a collective property of the
qubits. The 3-tangle is also an entanglement monotone \cite{Dur}.
Monotonicity for entanglement measure is a natural requirement.

The 3-tangle was extended to even $n$ qubits, and the extension was called
the $n$-tangle \cite{Wong}. Let the state $|\psi \rangle
=\sum_{i_{1}i_{2}...i_{n}}a_{i_{1}i_{2}...i_{n}}|i_{1}i_{2}...i_{n}\rangle $%
, where $i_{1}$, $i_{2}$, ..., $i_{n}$ $\epsilon $\ $\{0,1\}$. The $n$%
-tangle was defined as \cite{Wong}

\begin{eqnarray}
\tau _{1...n} &=&2|S|,  \notag \\
S &=&\sum (a_{\alpha _{1}...\alpha _{n}}a_{\beta _{1}...\beta _{n}}a_{\gamma
_{1}...\gamma _{n}}a_{\delta _{1}...\delta _{n}}  \notag \\
&&\times \epsilon _{\alpha _{1}\beta _{1}}\epsilon _{\alpha _{2}\beta
_{2}}...\epsilon _{\alpha _{n-1}\beta _{n-1}}  \notag \\
&&\times \epsilon _{\gamma _{1}\delta _{1}}\epsilon _{\gamma _{2}\delta
_{2}}...\epsilon _{\gamma _{n-1}\delta _{n-1}}\epsilon _{\alpha _{n}\gamma
_{n}}\epsilon _{\beta _{n}\delta _{n}}),  \notag \\
&&  \label{n-tangle-1}
\end{eqnarray}%
$\ $where $|c|$ is the modulus of the complex number $c$, $\alpha _{l}$, $%
\beta _{l}$, $\gamma _{l}$, and $\delta _{l}$\ $\epsilon \{0,1\}$, and 
\begin{equation}
\epsilon _{00}=\epsilon _{11}=0\text{ and }\epsilon _{01}=-\epsilon _{10}=1.
\label{cond}
\end{equation}

The $n$-tangle of even $n$ qubits is invariant under permutations of the
qubits, and is an entanglement monotone \cite{Wong}. In \cite{Wong},\ the $n$%
-tangle was proposed as a potential entanglement measure.

The generalized CKW monogamy inequality for $n$ qubits was given by \cite%
{Yu05, Verstraete06}

\begin{equation}
C_{12}^{2}+...+C_{1n}^{2}\leq C_{1(2...n)}^{2}\text{.}  \label{Monogamy-1}
\end{equation}%
Here $\rho _{12}=tr_{3...n}\rho _{12...n}$, i.e., $\rho _{12}$ is obtained
from the density matrix $\rho _{12...n}$ by tracing out over qubits 3, ...,
and $n$, and $C_{1(23...n)}^{2}=4\det \rho _{1}$, where $\rho
_{1}=tr_{23...n}\rho _{123...n}$. Note that $C_{1(2...n)}$\ can be called
the concurrence between qubit 1 and qubits $2,...,$ and $n$ if qubits 2,
..., and $n$ are regarded as a single object. The difference\ between the
two sides of CKW monogamy inequality in Eq. (\ref{Monogamy-1}) can be
considered as a natural generalization of the residual entanglement of three
qubits to $n$ qubits, and was denoted as \cite{Yu05}\ 
\begin{equation}
\tau _{1(2...n)}=C_{1(2...n)}^{2}-(C_{12}^{2}+...+C_{1n}^{2})\text{.}
\label{residual-1}
\end{equation}%
\ 

In this paper, we investigate the relationship between the $n$-tangle and
the residual entanglement for any even $n\geq 4$\ qubits. This paper is
organized as follows. In Sec. 2, we show that the $n$-tangle is the square
of the SLOCC polynomial invariant of degree 2. In Sec. 3, we address the
relationship between the $n$-tangle and the residual entanglement of $n$
qubits. In Sec. 4, we summarize our results and conclusions.

\section{The $n$-tangle is the square of the SLOCC polynomial invariant of
degree 2}

The SLOCC invariants can be used for SLOCC\ classification and the
entanglement measure \cite{Sudbery, Luque, Leifer, Levay, LDF06, LDF07a,
Osterloh-JMP}. For four qubits, four independent SLOCC\ polynomial
invariants: $H$, $L$, $M$, and $D_{xt}$ were given in \cite{Luque}, where $H$
is of degree 2,$\ L$ and $M$ are of degree 4, and $D_{xt}$ is of degree 6.
Very recently, for four and five qubits,\ SL invariants of degrees 2 (for
only four qubits), 4, 6, 8, 10, 12 were studied in \cite{Osterloh-JMP}. The
antilinear operators \textquotedblleft combs\textquotedblright , which are
invariant under $SL(2,C)$, were constructed in \cite{Osterloh-PRA}. The
geometry of four qubit invariants was investigated in \cite{Levay}. For any
even $n$ qubits, the SLOCC polynomial invariant of degree 2 was given in 
\cite{LDF07a}. The SLOCC\ invariant of degree 4 of odd $n$ qubits was
discussed in \cite{LDF06, LDF07a}. Note that there are no invariants of
degree 2 for odd $n$\ qubits \cite{Luque}.

\subsection{Reduction of the $n$-tangle}

The $n$-tangle in Eq. (\ref{n-tangle-1}) is quartic and the computation of
the coefficients takes $3\ast 2^{4n}$ multiplications. Denote by $\overline{%
\alpha _{i}}$ the complement of $\alpha _{i}$. That is, $\overline{\alpha
_{i}}=0$ when $\alpha _{i}=1$. Otherwise, $\overline{\alpha _{i}}=1$.
Further, let

\begin{eqnarray}
S_{0} &=&\sum_{\alpha _{1}...\alpha _{n-1}}(a_{\alpha _{1}...\alpha
_{n-1}0}a_{\overline{\alpha _{1}}...\overline{\alpha _{n-1}}1}  \notag \\
&&\times \epsilon _{\alpha _{1}\overline{\alpha _{1}}}\epsilon _{\alpha _{2}%
\overline{\alpha _{2}}}...\epsilon _{\alpha _{n-1}\overline{\alpha _{n-1}}}).
\label{reduce}
\end{eqnarray}%
Note that $S_{0}$ is of degree 2. Then, $S$ in Eq. (\ref{n-tangle-1}) can be
reduced to $S=2S_{0}^{2}$ (see (A) of Appendix A for the proof). This leads
to 
\begin{equation}
\tau _{1...n}=|2S_{0}|^{2}\text{.}  \label{n-tang-7}
\end{equation}

\subsection{The $n$-tangle is the square of the SLOCC polynomial invariant
of degree 2}

Let $|\psi \rangle =\sum_{i=0}^{2^{n}-1}a_{i}|i\rangle $ and $|\psi ^{\prime
}\rangle =\sum_{i=0}^{2^{n}-1}b_{i}|i\rangle $ be any states of $n$ qubits.
Two states $|\psi \rangle $ and $|\psi ^{\prime }\rangle $ are SLOCC\
equivalent if and only if there exist invertible local operators $\mathcal{A}%
_{1}$, $\mathcal{A}_{2}$, ..., $\mathcal{A}_{n}$ such that \cite{Dur} 
\begin{equation}
|\psi ^{\prime }\rangle =\underbrace{\mathcal{A}_{1}\otimes \mathcal{A}%
_{2}\otimes ...\otimes \mathcal{A}_{n}}_{n}|\psi \rangle .
\end{equation}%
The entanglement measure of the state $|\psi \rangle $ of even $n$ qubits
was proposed as \cite{LDF07a, LDF09b} 
\begin{equation}
\tau ^{\prime }(\psi )=\ 2\left\vert \mathcal{I}^{\ast }(a,n)\right\vert ,
\label{SLOCC-invariant}
\end{equation}%
where%
\begin{eqnarray}
&&\mathcal{I}^{\ast }(a,n)=\sum_{l=0}^{2^{n-2}-1}[(-1)^{N(l)}  \notag \\
&&\times (a_{2l}a_{(2^{n}-1)-2l}-a_{2l+1}a_{(2^{n}-2)-2l})].
\label{SLOCC-invari}
\end{eqnarray}%
Here we take $N(l)$ to be the number of the occurrences of \textquotedblleft 
$1$\textquotedblright\ in $l_{n-1}...l_{1}l_{0}$, which is a $n$-bit binary
representation of $l$, i.e., $l=l_{n-1}2^{n-1}+...+l_{1}2^{1}+l_{0}2^{0}$.
In \cite{LDF07a}, it was proven that if $|\psi \rangle $ and $|\psi ^{\prime
}\rangle $ are SLOCC\ equivalent then

\begin{equation}
\mathcal{I}^{\ast }(b,n)=\mathcal{I}^{\ast }(a,n)\det (\mathcal{A}%
_{1})...\det (\mathcal{A}_{n})\text{,}  \label{invariant}
\end{equation}%
\ where $\mathcal{I}^{\ast }(b,n)$ is obtained from $\mathcal{I}^{\ast
}(a,n) $\ by replacing $a$ in $\mathcal{I}^{\ast }(a,n)$ with $b$, and $%
\mathcal{I}^{\ast }(a,n)$\ was called the SLOCC polynomial invariant of
degree 2 of even $n$ qubits.

Note that $S_{0}$ is just $\mathcal{I}^{\ast }(a,n)$ (see (B) in Appendix A
for the proof). By virtue of Eqs. (\ref{n-tang-7}) and (\ref{SLOCC-invariant}%
),\ we have $\tau _{1...n}=(\tau ^{\prime }(\psi ))^{2}$. It then follows
from Eqs. (\ref{SLOCC-invariant}) and (\ref{def-SLOCC}) that

\begin{equation}
\tau _{1...n}=4\left\vert
\sum_{l=0}^{2^{n-1}-1}(-1)^{N(l)}a_{2l}a_{(2^{n}-1)-2l}\right\vert ^{2}.
\label{SLOCC6}
\end{equation}

In Eq. (\ref{SLOCC6}), computing the coefficients requires $(2^{n-1}+2)$
multiplications. The $n$-tangle\ $\tau _{1...n}$\ is not considered as the
SLOCC polynomial invariant of degree 4 though $\tau _{1...n}$ is quartic and
satisfies the equation $\tau _{1...n}(|\psi ^{\prime }\rangle )=\tau
_{1...n}(|\psi \rangle )\det (\mathcal{A}_{1})...\det (\mathcal{A}_{n})$.
However, the square root of the $n$-tangle is the SLOCC polynomial invariant
of degree 2. The square root of the $n$-tangle turns out to be $\tau
^{\prime }(\psi )$. Using the properties of $\tau ^{\prime }(\psi )$ \cite%
{LDF07a, LDF09b}, the square root is also an entanglement monotone, and
invariant under permutations of all the qubits.

\section{Relationship between the $n$-tangle and the residual entanglement}

\subsection{The $n$-tangle is not the residual entanglement for any even $%
n\geq 4$\ qubits.}

To illustrate the relationship between n-tangle and residual entanglement,
we consider the following examples. For the $n$-qubit state $\alpha
_{1}|0...1\rangle +\alpha _{2}|0...010\rangle +...+\alpha _{n}|10...0\rangle 
$, equality in Eq. (\ref{Monogamy-1}) holds \cite{Coffman, Ou}, i.e. the
residual entanglement $\tau _{1(2...n)}=0$. According to Eq. (\ref{SLOCC6}),
it is easy to see that the n-tangle $\tau _{1...n}=0$. It follows that $\tau
_{1...n}=\tau _{1(2...n)}$. This is particularly true for the $n$-qubit
state $|W\rangle $ \cite{Wong}. For the state $|GHZ\rangle =\frac{1}{\sqrt{2}%
}(|0\rangle ^{\otimes n}+|1\rangle ^{\otimes n})$, the residual entanglement$%
\ \tau _{1(2...n)}=1$ \cite{Ou}, and the n-tangle$\ \tau _{1...n}=1$ \cite%
{Wong}. Thus, $\tau _{1...n}=\tau _{1(2...n)}$ for the state $|GHZ\rangle $.
Here is another example which gives $\tau _{1...n}=\tau _{1(2...n)}=$ $%
4\left\vert \alpha \gamma \right\vert ^{2}$ for the state of four qubits: $%
\alpha |0011\rangle +\beta |0110\rangle +\gamma |1100\rangle $ by utilizing
Eq. (\ref{SLOCC6}).

One might wonder if the two generalizations, which are the $n$-tangle $\tau
_{1...n}$ and the residual entanglement $\tau _{1(2...n)}$, are equal.
However, this is not always the case as the following example will show.
Consider, for example, the $n$-qubit symmetric Dicke states with $l$
excitations ($1\leq l\leq (n-1)$) \cite{Stockton}

\begin{equation}
|l,n\rangle =\sum_{i}P_{i}|1_{1}1_{2}...1_{l}0_{l+1}...0_{n}\rangle ,
\label{Dicke}
\end{equation}%
where $\{P_{i}\}$ is the set of all the distinct permutations of the qubits.
For the Dicke state $|(n/2),n\rangle $ with $(n/2)$ excitations of any even $%
n\geq 4$ qubits, Eq. (\ref{SLOCC6}) yields the $n$-tangle $\tau _{1...n}=1$.
In this case, $\rho _{12}\tilde{\rho _{12}}$ has only three nonzero
eigenvalues $(\frac{n}{2(n-1)})^{2}$, $(\frac{n-2}{4(n-1)})^{2}$ (double).
We then get the concurrence $C_{12}^{2}=\frac{1}{(n-1)^{2}}$. The symmetry
of the Dicke state leads to $C_{1i}^{2}=C_{12}^{2}$, $i=3,...,n$.
Calculating $C_{1(2...n)}$ further gives $C_{1(2...n)}^{2}=\allowbreak 1$.
In light of Eq. (\ref{residual-1}), the residual entanglement $\tau
_{1(2...n)}=\frac{n-2}{n-1}$. It says that for the Dicke state $%
|(n/2),n\rangle $, the $n$-tangle $\tau _{1...n}$ is greater than the
residual entanglement $\tau _{1(2...n)}$ and the difference is given by $%
\frac{1}{n-1}$.

\subsection{ A necessary and sufficient condition for the vanishing of the
concurrence $C_{1(2...n)}$}

For the state $|\psi \rangle =\sum_{i=0}^{2^{n}-1}a_{i}|i\rangle $ of $n$
qubits, the concurrence $C_{1(2...n)}$ can be written as

\begin{equation}
C_{1(2...n)}^{2}=4\sum_{0\leq i<j\leq
2^{n-1}-1}|a_{i}a_{j+2^{n-1}}-a_{i+2^{n-1}}a_{j}|^{2}.  \label{new-concur}
\end{equation}%
The right hand side of Eq. (\ref{new-concur}) turns out to be the sum of
squared moduli (see Appendix B for the proof).

In view of Eq. (\ref{new-concur}), any $n$-qubit concurrence

\noindent $C_{1(2...n)}$ vanishes if and only if the state is a product of a
state of one qubit and a state of $(n-1)$ qubits, i.e., the state is of the
form $|\phi \rangle _{1}\otimes $ $|\varphi \rangle _{2...n}$ (see Appendix
B for the proof). This allows one to understand how the concurrence $%
C_{1(2...n)}$ measures the entanglement of a state. In other words, the
concurrence $C_{1(2...n)}$\ is always positive unless the state is a product
of a state of one qubit and a state of $(n-1)$ qubits.\ In particular, this
is true for any entangled state of any $n$ qubits. That is, there exist $i$
and $j$ with $0\leq i<j\leq 2^{n-1}-1$, such that $a_{i}a_{j+2^{n-1}}\neq
a_{i+2^{n-1}}a_{j}$. It is worthwhile pointing out that the $n$-tangle
vanishes for some entangled states \cite{LDF09b}.

\subsection{The concurrence $C_{1(2...n)}\geq $ the $n$-tangle $\protect\tau %
_{1...n}$}

A closer examination of Eqs. (\ref{new-concur}) and (\ref{SLOCC-invari})
reveals that for even $n$\ qubits, the concurrence $C_{1(2...n)}$ is equal
to or greater than the $n$-tangle $\tau _{1...n}$ (see Appendix B for the
proof). We immediately have the following corollaries:

(1). For any state $|\psi \rangle $ of even $n$ qubits, if the concurrence C
vanishes then, clearly, so does the n-tangle.

(2). If the n-tangle $\tau _{1...n}$ of even $n$ qubits is positive, then
the concurrence $C_{1(2...n)}$ is also positive. \ 

\subsection{The residual entanglement is a partial measure for product states%
}

In this section, we show that for product state $|\psi \rangle
_{1...l}\otimes |\phi \rangle _{(l+1)...n}$ of any $n$ qubits, where $|\psi
\rangle $ is the state of the first $l$ qubits, the residual entanglement $%
\tau _{1(2...n)}$ for the product state is reduced to the residual
entanglement $\tau _{1(2...l)}$ for the state $|\psi \rangle $. First we
observe that $\rho _{1}(|\psi \rangle \otimes |\phi \rangle \langle \psi
|\otimes \langle \phi |)\allowbreak =\rho _{1}(|\psi \rangle \langle \psi |)$%
. By the definition of the concurrence, 
\begin{equation}
C_{1(2...n)}(|\psi \rangle \otimes |\phi \rangle )=C_{1(2...l)}(|\psi
\rangle ).  \label{partial-3}
\end{equation}%
That is, the concurrence $C_{1(2...n)}$ for the product state $|\psi \rangle
\otimes |\phi \rangle $ is just the concurrence $C_{1(2...l)}$ for the state 
$|\psi \rangle $. It tells us that the concurrence $C_{1(2...n)}$ only
measures the entanglement of the state $|\psi \rangle $.

Likewise, the concurrence $C_{1k}$ for the state $|\psi \rangle
_{1...l}\otimes |\phi \rangle _{(l+1)...n}$ is just the concurrence $C_{1k}$%
\ for the state $|\psi \rangle _{1...l}$, $k=2,...,l$. Since qubits 1 and $k$
are not entangled, the concurrence $C_{1k}$ for the state $|\psi \rangle
_{1...l}\otimes |\phi \rangle _{(l+1)...n}\ $vanishes for $k>l$. This can be
seen as follows.\ After some algebra, we find $\rho _{1(l+1)}\tilde{\rho}%
_{1(l+1)}=cI$, where c is a constant. It implies\ that the concurrence $%
C_{1(l+1)}=0$. In a similar manner we can show that the concurrence $%
C_{1k}=0 $ for $k\geq (l+2)$.\ This leads to

\begin{eqnarray}
&&C_{1k}(|\psi \rangle _{1...l}\otimes |\phi \rangle _{(l+1)...n})  \notag \\
&=&\left\{ 
\begin{array}{rc}
C_{1k}(|\psi \rangle _{1...l}), & 2\leq k\leq l \\ 
0, & l<k\leq n\text{.}%
\end{array}%
\right.  \label{partial-2}
\end{eqnarray}

Eqs. (\ref{partial-3}) and (\ref{partial-2}) together with the definition of
the residual entanglement\ give\ 
\begin{equation}
\tau _{1(2...n)}(|\psi \rangle \otimes |\phi \rangle )=\tau
_{1(2...l)}(|\psi \rangle ).  \label{partial-1}
\end{equation}%
This shows that the residual entanglement $\tau _{1(2...n)}$ for the product
state $|\psi \rangle \otimes |\phi \rangle $ is reduced to the residual
entanglement $\tau _{1(2...l)}$ for the state $|\psi \rangle $. It tells us
that $\tau _{1(2...n)}$ only measures the residual entanglement of the state 
$|\psi \rangle $.

However, for the product state $|\psi \rangle _{1...l}\otimes |\phi \rangle
_{(l+1)...n}$ of even $n$ qubits, when $|\psi \rangle $\ is a state of even $%
n$\ qubits, the n-tangle is multiplicative. That is, $\tau _{12...n}(|\psi
\rangle \otimes |\phi \rangle )=\tau _{12...l}(|\psi \rangle )\times \tau
_{12...(n-l)}(|\phi \rangle )$ \cite{LDF09b}.

The following example shows that the residual entanglement $\tau _{1(2...n)}$%
\ is not the n-way entanglement measure.\ For the product state $(\frac{1}{%
\sqrt{2}}(|000\rangle +|111\rangle ))^{\otimes 2k}$, by Eq. (\ref{partial-1}%
), the residual entanglement $\tau _{1(2...(6k))}=1$. It is worth noting
that the $n$-tangle is not the n-way entanglement measure either \cite{Wong}.

\section{\protect\LARGE Conclusion}

In summary, we have shown that the $n$-tangle is the square of the SLOCC
polynomial invariant of degree 2. We have found that the two
generalizations, namely the $n$-tangle and the residual entanglement of $n$%
-qubits, are different for any even $n\geq 4$ qubits. We have also proven
that the concurrence $C_{1(2...n)}$ vanishes if and only if the state is a
product of a state of one qubit and a state of $(n-1)$ qubits. In other
words, the concurrence $C_{1(2...n)}$\ is always positive unless the state
is a product of a state of one qubit and a state of $(n-1)$ qubits.
Furthermore, we have argued that the concurrence $C_{1(2...n)}$ is equal to
or greater than the $n$-tangle, and that the residual entanglement is a
partial measure for product states of any $n$ qubits.

\section*{Appendix A The $n$-tangle is the square of the SLOCC polynomial
invariant.}

\setcounter{equation}{0} \renewcommand{\theequation}{A\arabic{equation}}

(A). Proof of $S=2S_{0}^{2}$

In view of Eq. (\ref{cond}), we only need to consider $\beta _{i}=\overline{%
\alpha _{i}}$, $\delta _{i}=\overline{\gamma _{i}}$, $i=1$, ..., $(n-1)$, $%
\gamma _{n}=\overline{\alpha _{n}}$, and $\delta _{n}=\overline{\beta _{n}}$%
. Thus, Eq. (\ref{n-tangle-1}) becomes

\begin{eqnarray}
S &=&\sum (a_{\alpha _{1}...\alpha _{n-1}\alpha _{n}}a_{\overline{\alpha _{1}%
}...\overline{\alpha _{n-1}}\beta _{n}}a_{\gamma _{1}...\gamma _{n-1}%
\overline{\alpha _{n}}}a_{\overline{\gamma _{1}}...\overline{\gamma _{n-1}}%
\overline{\beta _{n}}}  \notag \\
&&\times \epsilon _{\alpha _{1}\overline{\alpha _{1}}}\epsilon _{\alpha _{2}%
\overline{\alpha _{2}}}...\epsilon _{\alpha _{n-1}\overline{\alpha _{n-1}}} 
\notag \\
&&\times \epsilon _{\gamma _{1}\overline{\gamma _{1}}}\epsilon _{\gamma _{2}%
\overline{\gamma _{2}}}...\times \epsilon _{\gamma _{n-1}\overline{\gamma
_{n-1}}}\epsilon _{\alpha _{n}\overline{\alpha _{n}}}\epsilon _{\beta _{n}%
\overline{\beta _{n}}}).  \label{n-tangle-2}
\end{eqnarray}%
We distinguish two cases.\ 

\vspace{3mm}

Case 1. $\beta _{n}=\alpha _{n}$.

In this case, $\epsilon _{\alpha _{n}\overline{\alpha _{n}}}\epsilon _{\beta
_{n}\overline{\beta _{n}}}=1$. Let

\begin{eqnarray}
S^{\prime } &=&\sum_{\gamma _{1}...\gamma _{n-1}}(a_{\gamma _{1}...\gamma
_{n-1}\overline{\alpha _{n}}}a_{\overline{\gamma _{1}}...\overline{\gamma
_{n-1}}\overline{\alpha _{n}}}  \notag \\
&&\times \epsilon _{\gamma _{1}\overline{\gamma _{1}}}\epsilon _{\gamma _{2}%
\overline{\gamma _{2}}}...\epsilon _{\gamma _{n-1}\overline{\gamma _{n-1}}}).
\end{eqnarray}

Then, Eq. (\ref{n-tangle-2}) becomes

\begin{eqnarray}
S &=&\sum_{\alpha _{1}...\alpha _{n-1}\alpha _{n}}(a_{\alpha _{1}...\alpha
_{n-1}\alpha _{n}}a_{\overline{\alpha _{1}}...\overline{\alpha _{n-1}}\alpha
_{n}}  \notag \\
&&\times \epsilon _{\alpha _{1}\overline{\alpha _{1}}}\epsilon _{\alpha _{2}%
\overline{\alpha _{2}}}...\epsilon _{\alpha _{n-1}\overline{\alpha _{n-1}}%
}\times S^{\prime }).  \label{n-tangle-3}
\end{eqnarray}

To compute $S^{\prime }$, we assume that $\overline{\alpha _{n}}$ is fixed
in $S^{\prime }$.$\ $For each term

\noindent $t=a_{\gamma _{1}...\gamma _{n-1}\overline{\alpha _{n}}}a_{%
\overline{\gamma _{1}}...\overline{\gamma _{n-1}}\overline{\alpha _{n}}}
\times $ $\epsilon _{\gamma _{1}\overline{\gamma _{1}}}\epsilon _{\gamma _{2}%
\overline{\gamma _{2}}}...\epsilon _{\gamma _{n-1}\overline{\gamma _{n-1}}}$%
, $\ S^{\prime }$ has the term

\noindent $t^{\prime }=a_{\overline{\gamma _{1}}...\overline{\gamma _{n-1}}%
\overline{\alpha _{n}}}a_{\gamma _{1}...\gamma _{n-1}\overline{\alpha _{n}}%
}\times \epsilon _{\overline{\gamma _{1}}\gamma _{1}}\epsilon _{\overline{%
\gamma _{2}}\gamma _{2}}...\epsilon _{\overline{\gamma _{n-1}}\gamma _{n-1}}$%
. Note that $\epsilon _{\gamma _{l}\overline{\gamma _{l}}}=-\epsilon _{%
\overline{\gamma _{l}}\gamma _{l}}$, $l=1,...,n$. Thus, $t=-t^{\prime }$ and
so $S^{\prime }=0$. Hence, $S=0$.

Case 2. $\beta _{n}=\overline{\alpha _{n}}$.

In this case, $\epsilon _{\alpha _{n}\overline{\alpha _{n}}}\epsilon _{\beta
_{n}\overline{\beta _{n}}}=-1$. Eq. (\ref{n-tangle-2}) becomes

\begin{eqnarray}
S &=&-\sum_{\alpha _{1}...\alpha _{n}}[a_{\alpha _{1}...\alpha _{n}}a_{%
\overline{\alpha _{1}}...\overline{\alpha _{n}}}\times \epsilon _{\alpha _{1}%
\overline{\alpha _{1}}}...\epsilon _{\alpha _{n-1}\overline{\alpha _{n-1}}} 
\notag \\
&&\mathcal{\times }\sum_{\gamma _{1}...\gamma _{n-1}}(a_{\gamma
_{1}...\gamma _{n-1}\overline{\alpha _{n}}}a_{\overline{\gamma _{1}}...%
\overline{\gamma _{n-1}}\alpha _{n}}  \notag \\
&&\times \epsilon _{\gamma _{1}\overline{\gamma _{1}}}...\epsilon _{\gamma
_{n-1}\overline{\gamma _{n-1}}})].  \label{n-tangle-4}
\end{eqnarray}%
\ Let

\begin{equation}
S_{i}=\sum_{\alpha _{1}...\alpha _{n-1}}(a_{\alpha _{1}...\alpha _{n-1}i}a_{%
\overline{\alpha _{1}}...\overline{\alpha _{n-1}}\overline{\imath }}\times
\epsilon _{\alpha _{1}\overline{\alpha _{1}}}...\epsilon _{\alpha _{n-1}%
\overline{\alpha _{n-1}}}),  \label{SLOCC-1}
\end{equation}%
where $i=0$, $1$. Thus, 
\begin{equation}
S=-2S_{0}S_{1}.  \label{reduce-1}
\end{equation}

Next we verify that $S_{1}=-S_{0}$. By the condition in Eq. (\ref{cond}), $%
\epsilon _{\alpha _{i}\overline{\alpha _{i}}}=-\epsilon _{\overline{\alpha
_{i}}\alpha _{i}}$, $i=1$, ..., $n$. Then,

\begin{eqnarray}
S_{1} &=&\sum_{\alpha _{1}...\alpha _{n-1}}(a_{\alpha _{1}...\alpha
_{n-1}1}a_{\overline{\alpha _{1}}...\overline{\alpha _{n-1}}0}  \notag \\
&&\times \epsilon _{\alpha _{1}\overline{\alpha _{1}}}\epsilon _{\alpha _{2}%
\overline{\alpha _{2}}}...\epsilon _{\alpha _{n-1}\overline{\alpha _{n-1}}})
\notag \\
\ &=&-\sum_{\alpha _{1}...\alpha _{n-1}}(a_{\overline{\alpha _{1}}...%
\overline{\alpha _{n-1}}0}a_{\alpha _{1}...\alpha _{n-1}1}  \notag \\
&&\times \epsilon _{\overline{\alpha _{1}}\alpha _{1}}\epsilon _{\overline{%
\alpha _{2}}\alpha _{2}}...\epsilon _{\overline{\alpha _{n-1}}\alpha _{n-1}})
\notag \\
\ &=&-S_{0}.  \label{SLOCC-2}
\end{eqnarray}%
Together the latter two equations yield the desired result.

\vspace{3mm}

(B). Proof of $S_{0}=\mathcal{I}^{\ast }(a,n)$

We can rewrite $\mathcal{I}^{\ast }(a,n)$ as

\begin{equation}
\mathcal{I}^{\ast
}(a,n)=\sum_{l=0}^{2^{n-1}-1}(-1)^{N(l)}a_{2l}a_{(2^{n}-1)-2l}.
\label{def-SLOCC}
\end{equation}%
Let $l_{n-1}...l_{1}$ be the $(n-1)$-bit binary number of $l$. Then, it
follows from Eq. (\ref{def-SLOCC}) that

\begin{eqnarray}
\mathcal{I}^{\ast }(a,n)
&=&\sum_{l_{n-1}...l_{2}l_{1}}(-1)^{N(l)}a_{l_{n-1}...l_{1}0}a_{\overline{%
l_{n-1}}...\overline{l_{1}}1}  \notag \\
&=&\sum_{l_{n-1}...l_{2}l_{1}}(a_{l_{n-1}...l_{1}0}a_{\overline{l_{n-1}}...%
\overline{l_{1}}1}  \notag \\
&&\times \epsilon _{l_{1}\overline{l_{1}}}\epsilon _{l_{2}\overline{l_{2}}%
}...\epsilon _{l_{n-1}\overline{l_{n-1}}})  \notag \\
&=&S_{0}\text{.}  \label{SLOCC-5}
\end{eqnarray}

The second equality follows by noting that

$(-1)^{N(l)}=\epsilon _{l_{1}\overline{l_{1}}}\epsilon _{l_{2}\overline{l_{2}%
}}...\epsilon _{l_{n-1}\overline{l_{n-1}}}$.\ 

\section*{Appendix B. Concurrence $C_{1(2...n)}$}

\setcounter{equation}{0} \renewcommand{\theequation}{B\arabic{equation}}

\ 

\textit{Result 1. Let the state }$|\psi \rangle
=\sum_{i=0}^{2^{n}-1}a_{i}|i\rangle $\textit{\ be any state of any }$n$%
\textit{\ qubits. Then }%
\begin{equation}
C_{1(2...n)}^{2}=4\sum_{0\leq i<j\leq
2^{n-1}-1}|a_{i}a_{j+2^{n-1}}-a_{i+2^{n-1}}a_{j}|^{2}.  \label{concur_1}
\end{equation}

\textit{Proof.}\ By direct calculation we find

\noindent $\det \rho
_{1}=\sum_{i,j=0}^{2^{n-1}-1}a_{i}a_{j+2^{n-1}}(a_{i}^{\ast
}a_{j+2^{n-1}}^{\ast }-a_{i+2^{n-1}}^{\ast }a_{j}^{\ast })$, where $%
a_{i}^{\ast }$ is the complex conjugate of $a_{i}$. By switching $i$ and $j$%
, the term $a_{i}a_{j+2^{n-1}}(a_{i}^{\ast }a_{j+2^{n-1}}^{\ast
}-a_{i+2^{n-1}}^{\ast }a_{j}^{\ast })$ becomes $a_{j}a_{i+2^{n-1}}(a_{j}^{%
\ast }a_{i+2^{n-1}}^{\ast }-a_{j+2^{n-1}}^{\ast }a_{i}^{\ast })$. Then

\begin{eqnarray}
&&a_{i}a_{j+2^{n-1}}(a_{i}^{\ast }a_{j+2^{n-1}}^{\ast }-a_{i+2^{n-1}}^{\ast
}a_{j}^{\ast })  \notag \\
&+&a_{j}a_{i+2^{n-1}}(a_{j}^{\ast }a_{i+2^{n-1}}^{\ast }-a_{j+2^{n-1}}^{\ast
}a_{i}^{\ast })  \notag \\
&=&|a_{i}a_{j+2^{n-1}}-a_{i+2^{n-1}}a_{j}|^{2}.  \label{concur_2}
\end{eqnarray}%
When $i=j$, the right side of Eq. (\ref{concur_2}) vanishes. So, $\det \rho
_{1}=\sum_{0\leq i<j\leq
2^{n-1}-1}|a_{i}a_{j+2^{n-1}}-a_{i+2^{n-1}}a_{j}|^{2}$. Since $%
C_{1(2...n)}^{2}=4\det \rho _{1}$ by definition, the desired result follows.

\vspace{3mm}

\textit{Result 2. For the state $|\psi \rangle $ of any $n$ qubits,}

\noindent $C_{1(2...n)}=0$\textit{\ if and only if }$|\psi \rangle $\textit{%
\ is a product of a state of one qubit and a state of }$(n-1)$\textit{\
qubits, i.e., }$|\psi \rangle =|\phi \rangle _{1}\otimes $\textit{\ }$%
|\varphi \rangle _{2...n}$\textit{.}

\textit{Proof.} Let $|\psi \rangle
=\dsum\limits_{i=0}^{2^{n}-1}a_{i}|i\rangle $. It is assumed that $%
C_{1(2...n)}=0$. Hence, by Eq. (\ref{concur_1}),

\begin{equation}
a_{i}a_{j+2^{n-1}}=a_{i+2^{n-1}}a_{j},  \label{cond-1}
\end{equation}%
where $0\leq i<j\leq 2^{n-1}-1$. We distinguish two cases.

Case 1. $\dsum\limits_{i=0}^{2^{n-1}-1}|a_{i}|^{2}=0$. It is straightforward
to verify that $|\psi \rangle =|1\rangle _{1}\otimes
\dsum\limits_{j=0}^{2^{n-1}-1}a_{j+2^{n-1}}|j\rangle _{2...n}$.

Case 2. $\dsum\limits_{i=0}^{2^{n-1}-1}|a_{i}|^{2}\neq 0$. Without loss of
generality, assume that $a_{0}\neq 0$. Let $\alpha =\frac{a_{2^{n-1}}}{a_{0}}
$. Then,%
\begin{equation}
a_{2^{n-1}}=\alpha a_{0}\text{.}  \label{assume-1}
\end{equation}%
Letting $i=0$ in Eq. (\ref{cond-1}), we obtain

\begin{equation}
a_{0}a_{j+2^{n-1}}=a_{2^{n-1}}a_{j}\text{,}  \label{cond-2}
\end{equation}%
where $j=1,2,...,2^{n-1}-1$. Substituting Eq. (\ref{assume-1}) into Eq. (\ref%
{cond-2}), we see that 
\begin{equation}
a_{j+2^{n-1}}=\alpha a_{j}\text{,}  \label{cond-3}
\end{equation}%
where $j=1,2,...,2^{n-1}-1$. From Eqs. (\ref{assume-1}) and (\ref{cond-3}), $%
|\psi \rangle $ can be rewritten as $|\psi \rangle =(|0\rangle _{1}+\alpha
|1\rangle _{1})\otimes \dsum\limits_{j=0}^{2^{n-1}-1}a_{j}|j\rangle _{2...n}$%
.

Conversely, if $|\psi \rangle =|\phi \rangle _{1}\otimes $ $|\varphi \rangle
_{2...n}$, then it is readily verified that $C_{1(2...n)}=0$.

\vspace{3mm}

\textit{Result 3. For even $n$ qubits, the concurrence }

\noindent $C_{1(2...n)}$\textit{\ is equal to or greater than the n-tangle }$%
\tau _{1...n}$\textit{.}

\textit{Proof.} We rewrite Eq. (\ref{SLOCC-invari}) as 
\begin{eqnarray}
&&\mathcal{I}^{\ast }(a,n)=\sum_{k=0}^{2^{n-2}-1}[(-1)^{N(k)}  \notag \\
&&\times (a_{k}a_{2^{n}-1-k}-a_{2^{n-1}-1-k}a_{2^{n-1}+k})].  \label{SLOCC-3}
\end{eqnarray}

To prove this, we note that $\mathcal{I}^{\ast }(a,n)$ can be written as
(see \cite{LDF09b})

\begin{equation}
\mathcal{I}^{\ast }(a,n)=\sum_{k=0}^{2^{n-1}-1}(-1)^{N(k)}a_{k}a_{2^{n}-1-k}.
\label{SLOCC-4}
\end{equation}%
From Eq. (\ref{SLOCC-4}), 
\begin{eqnarray}
\mathcal{I}^{\ast }(a,n)
&=&\sum_{k=0}^{2^{n-2}-1}(-1)^{N(k)}a_{k}a_{2^{n}-1-k}  \notag \\
&+&\sum_{k=2^{n-2}}^{2^{n-1}-1}(-1)^{N(k)}a_{k}a_{2^{n}-1-k}.
\label{reduce-2}
\end{eqnarray}%
Let $k=2^{n-1}-1-i$, in which case $N(k)+N(i)=n-1$. Then, the second sum of
the above equation becomes $%
-\sum_{i=0}^{2^{n-2}-1}(-1)^{N(i)}a_{2^{n-1}-1-i}a_{2^{n-1}+i}$. Thus, Eq. (%
\ref{SLOCC-3}) holds.

For any $n$ qubits, we may write Eq. (\ref{concur_1}) as 
\begin{eqnarray}
&&C_{1(2...n)}^{2}  \notag \\
&=&4\biggl\{\sum_{\substack{ 0\leq i\leq 2^{n-2}-1  \\ i<j\leq 2^{n-1}-1  \\ %
j\neq 2^{n-1}-1-i}} |a_{i}a_{j+2^{n-1}}-a_{i+2^{n-1}}a_{j}|^{2}  \notag \\
&&+\sum_{2^{n-2}\leq i<j\leq 2^{n-1}-1}
|a_{i}a_{j+2^{n-1}}-a_{i+2^{n-1}}a_{j}|^{2}  \notag \\
&&+\sum_{i=0}^{2^{n-2}-1}
|a_{i}a_{2^{n}-1-i}-a_{2^{n-1}-1-i}a_{2^{n-1}+i}|^{2}\biggr\}.  \notag \\
&&  \label{concur_3}
\end{eqnarray}

For even $n$ qubits, from Eq. (\ref{SLOCC-3})\ it holds that 
\begin{equation}
\tau _{1...n}\leq 4\biggl[%
\sum_{k=0}^{2^{n-2}-1}|a_{k}a_{2^{n}-1-k}-a_{2^{n-1}-1-k}a_{2^{n-1}+k}|%
\biggr]^{2}.  \label{n-tangle-7}
\end{equation}

Let, for brevity,

\noindent $Z_{k}=|a_{k}a_{2^{n}-1-k}-a_{2^{n-1}-1-k}a_{2^{n-1}+k}|$ and $%
P(i,j)=a_{i}a_{j+2^{n-1}}-a_{i+2^{n-1}}a_{j}$. To show $C_{1(2...n)}^{2}\geq
\tau _{1...n}$, from Eqs. (\ref{concur_3}) and (\ref{n-tangle-7}), it is
enough to prove

\begin{eqnarray}
&&\sum_{\substack{ 0\leq i\leq 2^{n-2}-1  \\ i<j\leq 2^{n-1}-1  \\ j\neq
2^{n-1}-1-i}}|P(i,j)|^{2} +\sum_{2^{n-2}\leq i<j\leq 2^{n-1}-1}|P(i,j)|^{2} 
\notag \\
&\geq &2\sum_{0\leq k<m\leq 2^{n-2}-1}Z_{k}Z_{m}.  \label{ineq-0}
\end{eqnarray}

Observe that in Eq. (\ref{ineq-0}), the first, second, and third sums
contain $3\times 2^{n-3}(2^{n-2}-1)$ different terms $|P(i,j)|^{2}$, $%
2^{n-3}(2^{n-2}-1)$ different terms$\ |P(i,j)|^{2}$, and $2^{n-3}(2^{n-2}-1)$
different terms $Z_{k}Z_{m}$, respectively. Next we show that for each term $%
Z_{k}Z_{m}$ on the right side of Eq. (\ref{ineq-0}), there exist four
different corresponding terms $|P(i,j)|^{2}$ on the left side of Eq. (\ref%
{ineq-0})\ such that their sum is equal to or greater than $2Z_{k}Z_{m}$.

Given $Z_{k}Z_{m}$ with $0\leq k<m\leq 2^{n-2}-1$. We first choose two
different terms$\ $ $|P(k,2^{n-1}-1-m)|^{2}$ and $|P(m,2^{n-1}-1-k)|^{2}$
from the first sum in Eq. (\ref{ineq-0}). It is trivial that

\begin{eqnarray}
&&|P(k,2^{n-1}-1-m)|^{2}+|P(m,2^{n-1}-1-k)|^{2}  \notag \\
&\geq &2|P(k,2^{n-1}-1-m)||P(m,2^{n-1}-1-k)|.  \notag \\
&&  \label{ineq-1}
\end{eqnarray}
We then choose the term $|P(k,m)|^{2}$ from the first sum in Eq. (\ref%
{ineq-0}) and the term $|P(2^{n-1}-1-m,2^{n-1}-1-k)|^{2}$ from the second
sum in Eq. (\ref{ineq-0}). It is trivial that 
\begin{eqnarray}
&&|P(k,m)|^{2}+|P(2^{n-1}-1-m,2^{n-1}-1-k)|^{2}  \notag \\
&\geq &2|P(k,m)||P(2^{n-1}-1-m,2^{n-1}-1-k)|.  \notag \\
&&  \label{ineq-2}
\end{eqnarray}%
Now, using the fact that $|x|+|y|\geq |x-y|$, from Eqs. (\ref{ineq-1}) and (%
\ref{ineq-2}), we establish the inequality 
\begin{eqnarray}
&&|P(k,2^{n-1}-1-m)||P(m,2^{n-1}-1-k)|  \notag \\
&+&|P(k,m)||P(2^{n-1}-1-m,2^{n-1}-1-k)|  \notag \\
&\geq &Z_{k}Z_{m},  \label{ineq-3}
\end{eqnarray}%
and this implies the desired result Eq. (\ref{ineq-0}). This completes the
proof.

\end{document}